\documentclass{article}
\usepackage{spconf, amsmath, amssymb, graphicx}
\usepackage{pstricks, pstricks-add, pst-node, pst-tree}
\usepackage[bookmarks=true, bookmarksnumbered=true, bookmarksopen=true,
  unicode=true, colorlinks=true,
  pagebackref=true,
  linkcolor=blue,citecolor=blue,filecolor=blue,urlcolor=blue,
  pdfstartview=FitH
]{hyperref}
\usepackage{color}

\title{CONSTRAINED SPEAKER DIARIZATION OF \textsc{tv} SERIES\\BASED ON VISUAL PATTERNS}

\name{Xavier Bost, Georges Linar\`es\sthanks{This work was partially
    supported by the French National Research Agency (\textsc{anr})
    \textsc{contnomina} project (\textsc{anr}-07-240) and the Research
    Federation \textit{Agorantic}, Avignon University.}}

\address{LIA, University of Avignon, 339 chemin des Meinajari{\`e}s,
  84000 Avignon, France}
\begin{document}
\ninept
\maketitle

\begin{abstract}
  Speaker diarization, usually denoted as the ``who spoke when'' task,
  turns out to be particularly challenging when applied to fictional
  films, where many characters talk in various acoustic conditions
  (background music, sound effects...). Despite this acoustic
  variability, such movies exhibit specific visual patterns in the
  dialogue scenes. In this paper, we introduce a two-step method to
  achieve speaker diarization in \textsc{tv} series: a speaker
  diarization is first performed locally in the scenes detected as
  dialogues; then, the hypothesized local speakers are merged in a
  second agglomerative clustering process, with the constraint that
  speakers locally hypothesized to be distinct must not be assigned to
  the same cluster. The performances of our approach are compared to
  those obtained by standard speaker diarization tools applied to the
  same data.
\end{abstract}

\vspace{2mm}

\noindent \textcolor{red}{\textbf{Cite as:}\\X.~Bost \& G.~Linar\`es.\\\href{https://ieeexplore.ieee.org/document/7078606}{Constrained speaker diarization of TV series based on visual patterns.}\\2014 IEEE Spoken Language Technology Workshop (SLT).\\doi: \href{https://doi.org/10.1109/SLT.2014.7078606}{10.1109/SLT.2014.7078606}}

\section{Introduction}
\label{sec:intro}

Speaker diarization is defined as the task of assigning the utterances
contained in a spoken document to their respective speakers. For this
purpose, two steps are involved, either sequentially or in
conjunction: detection of change points between speakers; clustering
of the resulting spoken segments in order to assign to the same
speaker its own utterances. The clustering step is usually achieved in
an iterative hierarchical process, either by agglomerating the closest
spoken segments into the same cluster in a bottom-up strategy or by
splitting the whole stream of utterances into smaller clusters in a
top-down way.

This task is usually performed as an unsupervised one, in particular
without allowing any prior knowledge of the number of speakers. This
lack of information makes the stop condition of the hierarchical
clustering process quite critical.

Speaker diarization (\textsc{sd}) systems were first developed for
processing of audio-only streams in adverse --~but controlled~--
acoustic conditions, such as telephone conversations, broadcast news,
meetings...  Some recent works applied them to videos whose production
context is uncontrolled, facing difficulties due to content and
environment variabilities.

In~\cite{clement2011speaker}, the authors apply standard \textsc{sd}
tools to the audio source of various kinds of video documents. The
reported results exhibit a Diarization Error Rate (\textsc{der}) much
higher than for those classical application fields.  The most dramatic
decrease in performance is observed when the \textsc{sd} systems are
applied to cartoons and movie trailers: among the possible reasons
involved, the authors notice the high number of speakers implied in
these kinds of stream, as well as the high variability of the acoustic
environment (speech and music segments overlapping each other, sound
effects). Moreover, as in most of previous related works on
audiovisual \textsc{sd}, diarization problem is here addressed by
applying audio-only systems to the audio channel of videos, without
any integration of the video-related features that could help the
diarization system.
 
However, some recent works focus on multimodal approaches for
performing speaker segmentation of video streams:
in~\cite{Friedland2009}, the authors evaluate a method based on early
fusion of audio and video \textsc{gmm}s, and a classical
\textsc{bic}-based agglomerative process on the resulting two-channel
information stream.  This technique is evaluated on the \textsc{ami}
corpus~\cite{carletta2005} that consists of audiovisual recordings of
four participants playing roles in a meeting scenario.

In this paper, we are interested in diarization of \textsc{tv} series, as a
major basic part of a wider project aiming at automatic structuring
of fictional videos.

Applying a \textsc{sd} system to \textsc{tv} series, where the speaker
number is generally higher than in full-length movies, may thus be
expected to be quite challenging. Nevertheless, fictional films
exhibit formal regularities at a visual level. For instance, dialogue
scenes require that the ``180-degree'' convention be respected in
order to preserve the visual fluidity of the exchange: so that both
speakers seem to look at each other when they appear successively on
the screen, the first one must look right and the second one must look
left, resulting in keeping the two cameras along the same side of an
imaginary line connecting them. Such a rule results in a specific
visual pattern made of two alternating, recurring shots and highly
typical of a dialogue scene.

Relying on such patterns, we propose here to split the speaker
diarization process into two steps when applied to fictional films:
the first one consists in a local speaker diarization inside the
boundaries of the visually detected dialogue scenes; the next one
consists in clustering the local hypothesized speakers while
preventing speakers locally assumed to be distinct from being merged
into the same cluster and propagating this constraint at each
iteration of the process.

Such a two-step clustering process is somehow related to what is
denoted in~\cite{tran2011comparing} as the ``hybrid architecture'' in
the cross-show speaker diarization context. In cross-show \textsc{sd},
diarization is achieved on a set of shows originating from a same
source and containing possibly recurring speakers. The shows are first
processed independently, before the resulting hypothesized speakers
are clustered in a second stage. In~\cite{bendris2013unsupervised},
the authors make use of speaker diarization in conjunction with face
clustering to identify the persons involved in a debate video: the
best modality to identify a person is chosen and the identity
information acquired for an instance is propagated to its whole
cluster. Finally, speaker diarization has already been applied to
\textsc{tv} series, but as a mean, among other modalities, to segment
the whole video stream into homogeneous narrative
scenes. In~\cite{bredin2012segmentation}, the performances of
mono-modal and multi-modal approaches for the scene segmentation task
are evaluated and compared.

In this paper, rather than using speaker diarization to structure the
\textsc{tv} movie, we propose to use its structure, as hypothesized
from visual patterns, to improve the speaker diarization of such
contents. The way such visual patterns are extracted is described in
Section~\ref{sec:patterns}. The two steps of our speaker diarization
approach, as well as the acoustic features used, are described in
Section~\ref{sec:diar}. Experimental results are presented and
discussed in Section~\ref{sec:exp}.

\section{Visual patterns detection}
\label{sec:patterns}

The whole video stream can be regarded as a finite sequence of fixed
images (or video frames) displayed on the screen at a constant rate to
simulate motion continuity. As mentioned
in~\cite{koprinska2001temporal}, a shot is defined as ``an unbroken
sequence of frames taken from one camera''.

As noticed in Section~\ref{sec:intro}, because of technical narrative
constraints, recurring and alternating shots frequently occur in the
dialogue scenes of fictional movies, resulting in specific patterns.

In order to automatically extract such patterns, we then first need to
split the whole video stream into shots and compare them to detect the
recurring ones.

\subsection{Shot segmentation and detection of similar shots}
\label{ssec:shotsegsim}

Defined by the continuity of the images it contains, a shot can also
be defined, in a contrastive way, in opposition to the previous
one. Shot segmentation is thus classically performed by detecting the
transitions, either abrupt or gradual, between temporally contiguous
shots~(\cite{koprinska2001temporal}). Remaining marginal
in~\textsc{tv} series, gradual transitions are here discarded and only
abrupt ones (or cuts) are considered.

A cut between two contiguous shots is hypothesized if two temporally
adjacent images differ from each other beyond a given threshold
$\tau_1$. Similarly, the present shot and a past one are considered as
similar if the difference between the first image of the former and
the last image of the latter stays below another threshold $\tau_2$.

Both tasks, shot cut detection as well as shot similarity detection,
require that two images be compared. 3-dimension histograms of the
image pixel values in the \textsc{hsv} color space are used to
describe the image. However, two different images may share the same
color histogram, resulting in an irrelevant similarity: spatial
information about the color distribution on images is reintroduced by
splitting the whole image into 30 pixel blocks, each associated with
its own histogram; block-based comparison of the resulting local
histograms, as described in~\cite{koprinska2001temporal}, is then
performed. The similarity between two color histograms is measured by
their correlation.

The two thresholds $\tau_1$ and $\tau_2$ respectively needed to
achieve both tasks, shot cut detection and shot similarity detection,
are estimated by experiments on a development set.

\subsection{Shot patterns extraction}
\label{ssec:shotpatt}

Let $\Sigma = \{l_1, ..., l_m\}$ be a finite set of $m$ shot labels,
two shots sharing the same label if they are hypothesized as similar
as stated in the subsection~\ref{ssec:shotsegsim}.

The whole movie can then be described by a finite string $\mathbf{s} =
s_1s_2...s_k$ of shot labels, with each $s_i \in \Sigma$.

For any couple of shot labels $(l_1, l_2) \in \Sigma^2$, the following
regular expression $r(l_1, l_2)$ denotes a subset of the set of all
the possible shot label sequences $\Sigma^* = \bigcup_{n \geqslant 0}
\Sigma^n$~:

\begin{equation}
  r(l_1, l_2) = \Sigma^* l_1(l_2l_1)^+ \Sigma^*
  \label{eq:patt1}
\end{equation}

The set $\mathcal{L}(r(l_1, l_2))$ of strings denoted by such a
regular expression corresponds to all the shot label sequences
containing $l_2$ inserted between two occurrences of $l_1$ with a
possible repetition of the alternation $l_2l_1$, whatever be the
previous and following shot labels. This regular expression formalize
the intuition of the ``two-alternating-and-recurring-shots'' pattern
mentioned in section~\ref{sec:intro} and typical of dialogue scenes.

\begin{figure}[htb]
  \vspace{0.5cm}
  \begin{minipage}[b]{1.0\linewidth}
    \centering
    \centerline{\includegraphics[width=9cm]{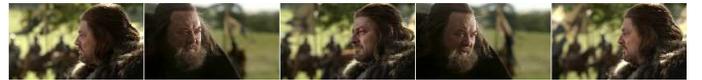}}
  \end{minipage}
  \caption{{\it Example of shot sequence $...l_1l_2l_1l_2l_1...$
      captured by the regular expression~\ref{eq:patt1} for two shot
      labels $l_1$ and $l_2$.}}
  \vspace{0.5cm}
  \label{seq1}
\end{figure}

Figure~\ref{seq1} shows a sequence of shots captured by the regular
expression~\ref{eq:patt1}, as well as it illustrates the
``180-degree'' rule mentioned in section~\ref{sec:intro}.

For a given movie described by a sequence $\mathbf{s} = s_1s_2...s_k$
of shot labels, a set $\mathcal{P}(\mathbf{s}) \subseteq \Sigma^2$ of
shot patterns is extracted by considering all the couples of shot
labels $(l_1, l_2) \in \Sigma^2$ such that $\mathbf{s} \in
\mathcal{L}(r(l_1, l_2))$:

\begin{equation}
  \mathcal{P}(\mathbf{s}) = \{(l_1, l_2) \ | \ \mathbf{s} \in
  \mathcal{L}(r(l_1, l_2))\}
\end{equation}

In other words, $\mathcal{P}(\mathbf{s})$ contains all the label pairs
which occur as recurring subsequences of the form showed on
Figure~\ref{seq1} in the whole movie sequence $\textbf{s}$.

The set of utterances $\mathbf{u}(l_1, l_2)$ covered by the pattern
$(l_1, l_2)$ are then all these which occur whenever the two shots
alternate with each other according to rule~\ref{eq:patt1}.

In order to increase the coverage of the patterns included in
$\mathcal{P}(\mathbf{s})$ and reduce their sparsity, two extensions or
the condition~\ref{eq:patt1} are introduced.

\begin{figure}[htb]
  \vspace{0.5cm}
  \begin{minipage}[b]{1.0\linewidth}
    \centering
    \centerline{\includegraphics[width=9cm]{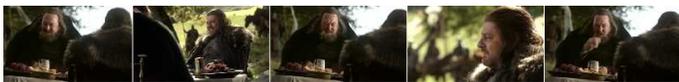}}
  \end{minipage}
  \caption{{\it Shot sequence $...l_1l_2l_1l_3l_1...$ at the boundary
      of two adjacent patterns $(l_1, l_2)$ and $(l_1, l_3)$ with one
      shot in common.}}
  \vspace{0.5cm}
  \label{seq2}
\end{figure}

\begin{enumerate}

\item In addition to rule~\ref{eq:patt1}, isolated expressions of the two
  alternating shots of the form $(l_1l_2 | l_2l_1)^+$ are taken into
  account, increasing the total amount of speech captured by the patterns.

\item The number of patterns is reduced while the average pattern
  coverage is increased by iteratively merging in a new pattern two
  patterns $(l_1, l_2)$ and $(l_1, l_3)$ with at least one label in
  common. As showed in Figure~\ref{seq2}, such situations frequently
  occur during dialogues when one of the speakers (here the one
  appearing on the shots $l_2$ and $l_3$) is alternatively filmed from
  two distinct cameras. The resulting pattern gather all the utterances
  $\mathbf{u}(l_1, l_2)$ and $\mathbf{u}(l_1, l_3)$ covered by the
  merged patterns.

\end{enumerate}

Table~\ref{pattstat} reports the total coverage of the patterns
extracted from the movies of our corpus (described in
subsection~\ref{ssec:corpus}), expressed as the ratio between the
amount of speech covered by the patterns and the total amount of
speech. The average duration of the speech covered by each pattern is
also indicated, as well as the average number of speakers by
pattern. These data are both computed by applying the basic version of
the regular expression $r$, as given in equation~\ref{eq:patt1}, and
by using the extended expression of $r$.

\begin{table}[h]
  \caption{\label{pattstat}{\it Shot patterns and speech: statistical data}}
  \vspace{2mm}
  \centering
  \begin{tabular}{|c||c|c|c|}
    \hline
    & \textbf{coverage (\%)} & \textbf{spch/patt (s.)} & \textbf{\# of
      spk/patt} \\
    \hline
    $r$ & 49.51 & 11.07 & 1.77 \\
    \hline
    ext. $r$ & \textbf{51.99} & \textbf{20.90} & \textbf{1.86} \\
    \hline
  \end{tabular}
\end{table}

As indicated in Table~\ref{pattstat}, the extracted visual patterns
cover in average a bit more than half (51.99\%) of the total amount of
speech contained in the \textsc{tv} movies of our corpus.

69.85\% of the patterns contain 2 speakers, 8.09\% three and 22.06\%
only one. However, most of these one-speaker patterns correspond to
short scenes, where the probability that one of the speakers remains
silent increases.

\begin{figure}[htb]
  \begin{minipage}[b]{1.0\linewidth}
    \centering
    \centerline{\includegraphics[width=9cm]{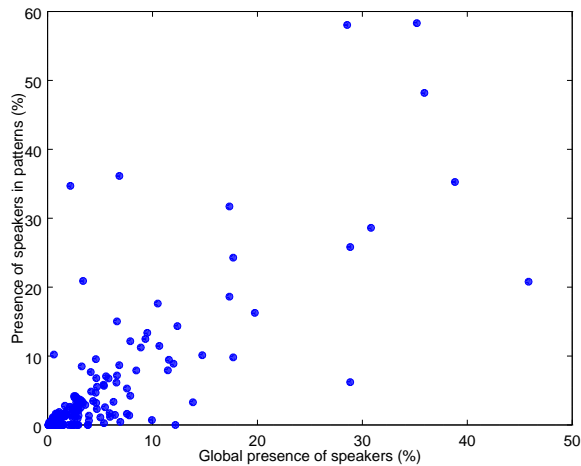}}
  \end{minipage}
  \caption{{\it Global \textit{vs} pattern presence of each speaker}}
  \label{repr}
\end{figure}

Figure~\ref{repr} shows the ability of such visual patterns to capture
the main characters of a narrative movie: 97.96\% of the characters
speaking at least 5\% of the time are involved in such patterns.

\section{Speaker diarization}
\label{sec:diar}

Speaker diarization is performed in two steps: speaker diarization is
first achieved locally by clustering the set of utterances
$\textbf{u}(l_1, l_2)$ covered by the pattern $(l_1, l_2)$; in a
second stage, the locally hypothesized speakers are clustered in order
to merge recurring speakers.

\subsection{Acoustic features}
\label{ssec:acfeatures}

Easily available, the subtitles of the movie are here used as an way
to estimate the boundaries of the corresponding speech segments. As an
exact transcription of the speech uttered, the subtitles temporally
match it, despite a slight and variable latency before they are
displayed on the screen and after they disappear. When the latency was
too large, the subtitle boundaries were manually adjusted.

Moreover, a subtitle generally corresponds to a spoken segment uttered
by a single speaker; on the remaining ones that cover two speech
turns, the boundaries of each utterance are indicated, allowing to
split the whole subtitle into two shorter ones.

The detection of change points between the possible audio sources, as
a prerequisite of most of the diarization systems, is thus here
avoided, allowing us to focus on the clustering process.

The acoustic parameterization of the resulting spoken segments is
achieved by extracting 19 cepstral coefficients plus energy, completed
by their first and second derivatives.

As a state of the art approach in the speaker verification field,
i-vectors are used to retain the relevant acoustic information from
each spoken segment (\cite{dehak2011front}). I-vectors are extracted
by using a 512-components \textsc{gmm/ubm} and a total variability
matrix trained on a development set.

The initial set of instances to cluster is then made of 60-dimension
normalized i-vectors, each corresponding to a speech segment uttered
by a single speaker.

\subsection{Agglomerative local clustering}
\label{ssec:locclust}

A first step of agglomerative clustering is processed within each
local dialogue scene as hypothesized by the use of the visual patterns
described in subsection~\ref{ssec:shotpatt}.

For the set of utterances $\mathbf{u}(l_1, l_2)$ covered by the
pattern $(l_1, l_2)$, the bottom-up clustering algorithm relies on the following:

\begin{itemize}

  \item The Mahalanobis distance is chosen as a similarity measure
    between the i-vectors corresponding to the spoken segments,
    resulting in a matrix $M$ of similarity between the utterances
    contained in $\mathbf{u}(l_1, l_2)$.
    
    The covariance matrix used to compute the Mahalanobis distance is
    the within class covariance matrix of the training set, as
    mentioned in~\cite{bousquet2011intersession} and computed as
    follows:

    \begin{equation}
      W = \frac{1}{n} \sum_{s=1}^S \sum_{i=1} ^{n_s} (\mathbf{u_i^s} -
      \overline{\mathbf{u_s}}) (\mathbf{u_i^s} -
      \overline{\mathbf{u_s}})^T
    \end{equation}

    where $n$ denotes the number of spoken segments of the training
    set, $S$ the number of speakers and $n_s$ the number of segments
    uttered by the speaker $s$; $\overline{\mathbf{u_s}}$ is the mean
    of the i-vectors corresponding to utterances of speaker $s$ and
    $\mathbf{u_i^s}$ denotes the i-vector corresponding to the $i-$th
    utterance of speaker $s$.

  \item The Ward's aggregation criterion is used during the
    agglomeration process to estimate the distance $\Delta I(c, c')$
    between the clusters $c$ and $c'$; it is computed as follows:

    \begin{equation}
      \Delta I(c, c') = \frac{m_c m_{c'}}{m_c + m_{c'}} d^2 (g_c,
      g_{c'})
      \label{eq:ward}
    \end{equation}

    where $m_c$ and $m_{c'}$ are the respective mass of the two
    clusters, $g_c$ and $g_{c'}$ their respective mass centers and
    $d(g_c, g_{c'})$ the distance between the mass centers.

  \item Finally, the Silhouette method is used to cut the dendogram
    resulting of the clustering process and obtain the final partition
    of the spoken segments. Described
    in~\cite{rousseeuw1987silhouettes}, the Silhouette method allows
    to automatically choose a convenient partition of the instance set
    by evaluating the quality of a each possible partition resulting
    from the clustering process. For a given partition, if instances
    appear closer to another cluster than to their own, the
    quality measure tends to decrease, and to increase if the instances
    are appropriately assigned to their respective clusters.

\end{itemize}

\subsection{Constrained global clustering}
\label{ssec:globclust}

Once the speaker diarization is performed inside each dialogue scene,
a second stage of clustering is performed in order to merge the
recurring speakers.

The set of segments locally clustered as uttered by the same speaker
are extracted in order to be modelled by a speaker normalized i-vector
$\mathbf{s_i}$ of 60 components.

The global clustering of the resulting set is processed in the same
way than the local one, using Mahalanobis distance based on the $W$
covariance matrix, Ward's aggregation criterion and the Silhouette
method to extract the final partition of speakers.

However, this second step is guided, at each agglomeration step, by
the structural information given by the visual segmentation of the
movie into dialogue scenes as described in Section~\ref{sec:patterns}:
the global clustering step has to prevent speakers locally
hypothesized to be distinct from being assigned to the same cluster
during the iterative agglomeration process.

The integration of such a constraint in the bottom-up clustering
algorithm is achieved in the following way:

\begin{itemize}

  \item In the initial matrix $M$ of the distances between the
    i-vectors corresponding to the locally hypothesized speakers, the
    distance $d(\mathbf{s}, \mathbf{s'})$ between two instances
    $\mathbf{s}$ and $\mathbf{s'}$ is set to $+\infty$ if the
    corresponding two speakers appear together in the same dialogue
    scene:

    \begin{equation}
      d(\mathbf{s}, \mathbf{s'}) = +\infty \Leftrightarrow \exists
      (l_1, l_2), \ \mathbf{u}(\mathbf{s}) \cup
      \mathbf{u}(\mathbf{s'}) \subseteq \mathbf{u}(l_1, l_2)
      \label{eq:rl1}
    \end{equation}

    where $(l_1, l_2)$ denotes a dialogue pattern, $\mathbf{u}(l_1,
    l_2)$, the set of utterances covered by the pattern $(l_1, l_2)$
    and $\mathbf{u}(\mathbf{s})$ the set of utterances assigned to the
    speaker $\mathbf{s}$ during the local clustering step.
    
  \item The distance $\Delta I(c, c')$ between the clusters $c$ and
    $c'$ is set to $+\infty$ if at least one instance of the first cluster
    is located at an infinite distance from an instance of the second
    one:

    \begin{equation}
      \Delta I(c, c') = +\infty \Leftrightarrow \exists (\mathbf{s},
      \mathbf{s'}) \in c \times c', \ d(\mathbf{s}, \mathbf{s'}) =
      +\infty
      \label{eq:rl2}
    \end{equation}

    where $\mathbf{s}$ and $\mathbf{s'}$ denote i-vectors
    corresponding to hypothesized speakers.

\end{itemize}

The application of rules~\ref{eq:rl1} and~\ref{eq:rl2} prevents two
distinct speakers from being clustered when choosing at each iteration
of the agglomerative process the two closest instances to merge.

\begin{figure}[htb]
  \begin{minipage}[b]{1.0\linewidth}
    \centering
    \begin{pspicture}(8, 2)
      \psline[linewidth=.1]{->}(4, 1)(4.5, 1)
      \pscircle[linecolor=gray](0.5, 0.5){.3}
      \rput(0.5, 0.5){\textcolor{gray}{$s_{21}$}}
      \pscircle(1, 1.5){.3}
      \rput(1, 1.5){$s_{11}$}
      \pscircle[linecolor=gray](3, 0.5){.3}
      \rput(3, 0.5){\textcolor{gray}{$s_{22}$}}
      \pscircle(2.5, 1.5){.3}
      \rput(2.5, 1.5){$s_{12}$}
      \pscurve[linecolor=gray]{-}(0.8, 0.5)(1.75, 0.3)(2.7, 0.5)
      \pscurve{-}(1.3, 1.5)(1.75, 1.55)(2.2, 1.5)
      \psline[linecolor=gray]{-}(0.8, 0.5)(2.2, 1.5)
      \psline[linecolor=gray]{-}(1.3, 1.5)(2.7, 0.5)
      \psframe[framearc=.2, linestyle = dotted](0, 0)(1.5, 2)
      \psframe[framearc=.2, linestyle = dotted](2, 0)(3.5, 2)
      \pscircle(6, 1.3){.65}
      \pscircle[linestyle=dashed](5.67, 1.3){.3}
      \rput(5.67, 1.3){$s_{11}$}
      \pscircle[linestyle=dashed](6.33, 1.3){.3}
      \rput(6.33, 1.3){$s_{12}$}
      \pscircle(5, 0.3){.3}
      \rput(5, 0.3){$s_{21}$}
      \pscircle(7, 0.3){.3}
      \rput(7, 0.3){$s_{22}$}
      \pscurve{-}(5.3, 0.3)(6, 0.2)(6.7, 0.3)
    \end{pspicture}
  \end{minipage}
  \caption{{\it First iteration of constrained global clustering}}
  \vspace{0.5cm}
  \label{clust}
\end{figure}
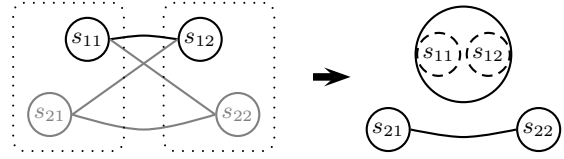

Figure~\ref{clust} illustrates both the application of these rules at
the initial step of the agglomerative process and how this
``different-speakers'' property is inherited by the newly created
cluster. Local dialogue scenes are surrounded by dotted rectangles;
each node $s_{ij}$ represents the $i-$th speaker hypothesized in the
$j-$th dialogue; the edges between two nodes represent their distance;
the absence of edge between two nodes corresponds to an infinite
distance. Merging the two closest nodes $s_{11}$ and $s_{12}$ results
in an isolated cluster $s_{11}s_{12}$ that inherits both from the
distinction between the two speakers of the first scene and from the
distinction between those of the second one: the hypothesized
recurring speaker in the two scenes has indeed to be different from
both the speakers he is respectively talking to.

Such a ``different-speaker'' property, as propagated at each step of
the agglomerative process, is expected to prevent the speakers
involved in a same dialogue to be prematurely clustered: the
background music of a dialogue may for instance hide the inter-speaker
variability and cause such an early clustering.

Moreover, the main consequence of respecting such a constraint is to
block the clustering process before assigning all the instances to the
same cluster. In the small example of Figure~\ref{clust}, only one
more step of the agglomerative process could be achieved, by
clustering $s_{21}$ and $s_{22}$: the narrative structure (two
dialogues with two speakers each) remains indeed compatible with such
a clustering. The resulting dendogram is then split into two distinct
trees.

\begin{figure}[htb]
  \begin{minipage}[b]{1.0\linewidth}
    \centering
    \centerline{\includegraphics[width=8.5cm]{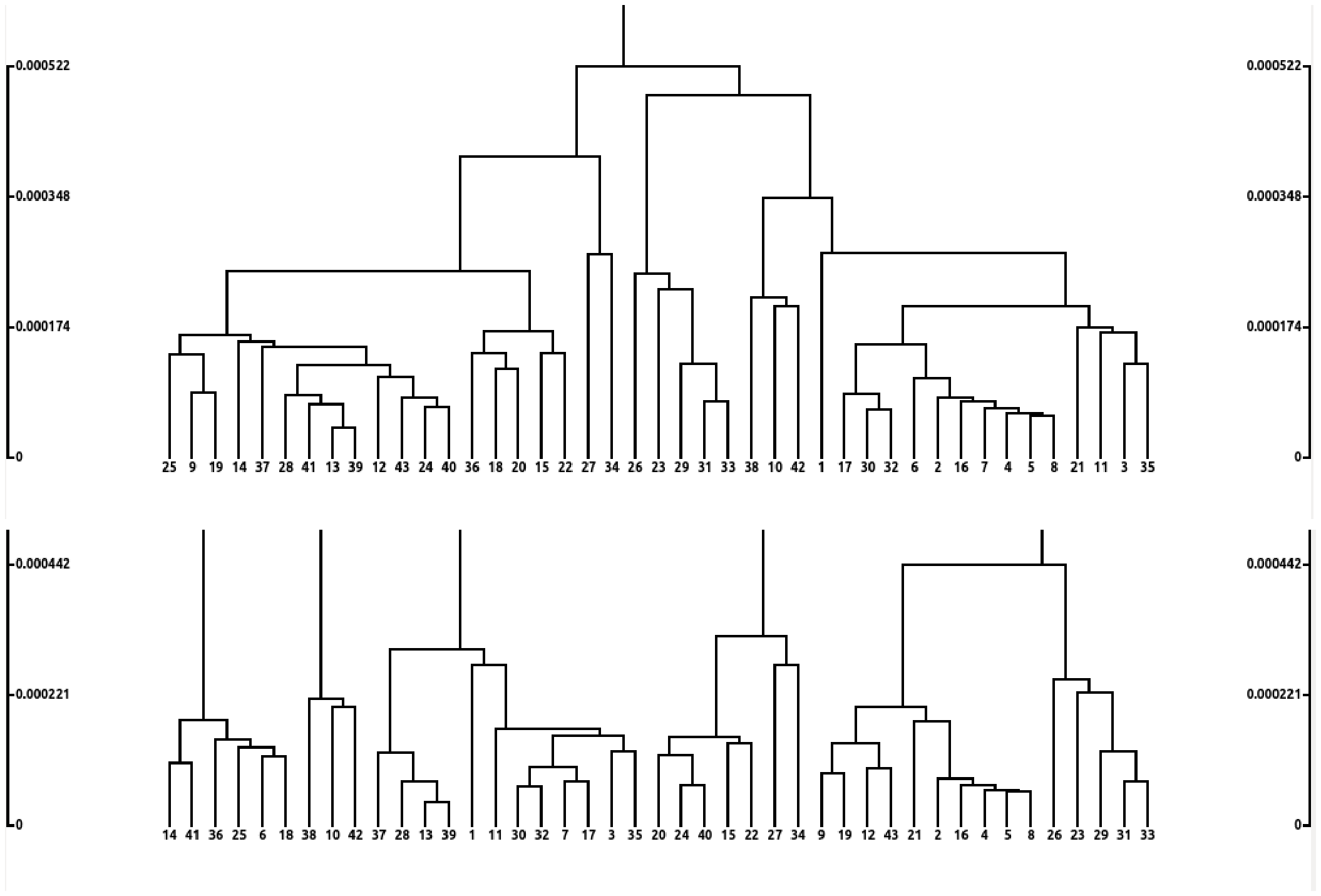}}
  \end{minipage}
  \caption{{\it Dendograms obtained by agglomerative clustering on
      local speaker hypotheses, unconstrained (top);
      constrained (bottom)}}
  \label{dendo}
\end{figure}

Figure~\ref{dendo} shows dendograms corresponding to agglomerative
clustering of local speakers. The one figuring on top is obtained in a
classical way, but may be difficult to cut automatically to extract
the best partition of the instances. The bottom part of the figure,
obtained with the same data by integrating the ``different-speakers''
property to the clustering process, shows five trees corresponding to
five incompatible groups of speakers; each one is made of a group of
narratively consistent speakers, with possibly many occurrences of the
same one.

Each of these remaining trees of compatible speakers is finally cut
using the Silhouette method described in
subsection~\ref{ssec:locclust} and the final partition of the instance
set is obtained by the union of the partitions obtained for each tree.

However, this constrained global clustering step remains dependent of
the outputs of the local one. If a single speaker is wrongly split
into two clusters during the local clustering step, the two resulting
utterance groups will never be merged during a global clustering
embedding the ``different-speakers'' property. Nevertheless, even
during an unconstrained clustering process, such groups would be
merged lately, possibly after the best partition is reached.

\section{Experiments and results}
\label{sec:exp}

\subsection{Corpus}
\label{ssec:corpus}

For experimental purpose, we acquired the first seasons of three
\textsc{tv} series: \textit{Breaking Bad} (abbreviated \textit{bb}),
\textit{Game of Thrones} (\textit{got}), and \textit{House of Cards}
(\textit{hoc}). We manually annotated three episodes of each series by
indicating shot cuts, similar shots, speech segments as well as the
corresponding speakers.

The total amount of speech in these nine episodes represents a bit
more than three hours (3:12).

A subset of six episodes (denoted \textsc{dev}) was used for
development purpose, the remaining three ones (denoted \textsc{test})
being used for test purpose.

\subsection{Shot cut and shot similarity detection}
\label{ssec:resshot}

The evaluation of shot cut detection relies on a classical F1-score
(\cite{boreczky1996comparison}) based on recall (\% of retrieved cuts
among the relevant ones) and precision (\% of relevant cuts among the
retrieved ones). For the shot similarity detection task, an analogous
F1-score is used: for each shot, the list of shots hypothesized as
similar to the current one is compared to the reference list of
similar shots; if both lists intersect in a non-empty set, the shot is
considered as correctly paired with its list. Results on \textsc{dev}
and \textsc{test} sets are reported in Table~\ref{resshot}.

\begin{table}[h]
  \caption{\label{resshot}{\it Results obtained for shot cut and shot similarity detection}}
  \vspace{2mm}
  \centering
  \begin{tabular}{|c|c||ccc|}
    \hline
    & \textbf{shot cut} & \multicolumn{3}{|c|}{\textbf{shot sim}} \\
    \cline{2-5}
    & F1-score & precision & recall & F1-score \\
    \hline
    \textit{bb-1} & 0.93 & 0.88 & 0.81 & 0.84 \\
    \hline
    \textit{bb-2} & 0.99 & 0.90 & 0.83 & 0.86 \\
    \hline
    \textit{got-1} & 0.97 & 0.88 & 0.84 & 0.86 \\
    \hline
    \textit{got-2} & 0.98 & 0.89 & 0.90 & 0.90 \\
    \hline
    \textit{hoc-1} & 0.99 & 0.91 & 0.92 & 0.92 \\
    \hline
    \textit{hoc-2} & 0.98 & 0.93 & 0.97 & 0.95 \\
    \hline
    \hline
    avg. \textsc{dev} & \textbf{0.97} & 0.90 & 0.88 & \textbf{0.89} \\
    \hline
    \hline
    \textit{bb-3} & 0.98 & 0.83 & 0.84 & 0.83 \\
    \hline
    \textit{got-3} & 0.99 & 0.92 & 0.89 & 0.91 \\
    \hline
    \textit{hoc-3} & 0.99 & 0.98 & 0.96 & 0.97 \\
    \hline
    \hline
    avg. \textsc{test} & \textbf{0.99} & 0.91 & 0.90 & \textbf{0.90} \\
    \hline
  \end{tabular}
\end{table}

The results obtained in both the image processing tasks, particularly
for the shot similarity detection one (F1-score amounting to 0.90) are
thus expected to provide a firm base for guiding speaker diarization
of narrative movies. Precision is slightly more important than recall,
resulting in some missed similarities between shots but with fewer
false positives. As a result, the dialogue patterns are slightly less
covering when based on automatic similarity detection (49.70\% of the
part-of-speech \textit{vs} 51.99\% when shot similarity is manually
indicated) but appear highly reliable.

\subsection{Local speaker diarization}
\label{ssec:resloc}

The \textsc{der} used to evaluate the local clustering step is
computed independently in each episode dialogue before averaging the
obtained scores according to each dialogue duration
(\textit{single-show} \textsc{der}, as mentioned
in~\cite{rouvier2013open}). The results are reported in
Table~\ref{resloc}, when using both the reference (denoted
\textit{ref.}) and the automatically detected (denoted \textit{auto.})
similar shots. For the sake of comparison, agglomerative clustering
(denoted \textsc{ac}), is compared to a ``naive method'' relying on a
strong assumption of synchronization between the audio and video
streams: clustering of local utterances is performed by assigning each
spoken segment the label of the current shot, assuming the two
alternating shots match exactly the speaker turns.

\begin{table}[h]
  \caption{\label{resloc}{\it Single-show \textsc{der} by episode
      obtained for the local diarization step}}
  \vspace{2mm}
  \centering
  \begin{tabular}{|c|cc||cc|}
    \hline
    & \multicolumn{2}{|c||}{\textbf{input auto.}} &
    \multicolumn{2}{|c|}{\textbf{input ref.}} \\
    \cline{2-5}
    & \textit{naive} & \textsc{ac} & \textit{naive} & \textsc{ac} \\
    \hline
    \hline
    \textit{bb-1} & 30.26 & \textbf{19.11} & 22.81 & 21.00 \\
    \hline
    \textit{bb-2} & 22.06 & 22.51 & 19.78 & \textbf{19.14} \\
    \hline
    \textit{got-1} & 22.16 & 23.70 & 19.46 & \textbf{15.78} \\
    \hline
    \textit{got-2} & 26.19 & 18.78 & 22.80 & \textbf{16.61} \\
    \hline
    \textit{hoc-1} & 17.23 & 13.36 & 16.31 & \textbf{11.84} \\
    \hline
    \textit{hoc-2} & 30.66 & \textbf{18.18} & 31.87 & 19.12 \\
    \hline
    \hline
    avg. \textsc{dev} & 24.76 & 19.27 & 22.17 & \textbf{17.25} \\
    \hline
    \hline
    \textit{bb-3} & 40.45 & 21.15 & 24.31 & \textbf{12.15} \\
    \hline
    \textit{got-3} & 33.45 & 17.43 & 35.43 & \textbf{12.80} \\
    \hline
    \textit{hoc-3} & 24.44 & 12.83 & 22.95 & \textbf{12.82} \\
    \hline
    \hline
    avg. \textsc{test} & 32.78 & 17.14 & 27.56 & \textbf{12.59} \\
    \hline
  \end{tabular}
\end{table}

The results obtained by performing an audio-based clustering of the
utterances of each dialogue scene appear better than those obtained by
applying the naive image-based method.

Moreover, the automation of the previous step, though slightly
degrading performances in speaker diarization, does not really impact
it, which confirms the reliability of the visual modality.

\subsection{Global speaker diarization}
\label{ssec:resglob}

Table~\ref{resglob} reports the results obtained during the clustering
of the local speakers, achieving the second step of the speaker
diarization process.

\begin{table}[h]
  \caption{\label{resglob}{\it \textsc{der} obtained for the global
      diarization step}}
  \vspace{2mm}
  \centering
  \begin{tabular}{|c|cc||cc||cc|}
    \hline
    & \multicolumn{2}{|c||}{\textbf{input auto.}} &
    \multicolumn{2}{|c||}{\textbf{input ref.}} & \multicolumn{2}{|c|}{\textbf{spch ref.}}\\
    \cline{2-7}
    & \textit{2S} & \textit{cst. 2S} & \textit{2S} & \textit{cst. 2S}
    & \textsc{lia} & \textsc{lium}\\
    \hline
    \hline
    \textit{bb-1} & 51.36 & 56.00 & 52.66 & \textbf{48.10} & 72.06 & 67.21 \\
    \hline
    \textit{bb-2} & \textbf{41.83} & 65.07 & 58.76 & 49.49 & 77.03 & 76.79 \\
    \hline
    \textit{got-1} & 70.13 & \textbf{52.79} & 70.67 & 53.87 & 65.57 & 58.49 \\
    \hline
    \textit{got-2} & 67.28 & \textbf{38.85} & 70.32 & 41.24 & 65.29 & 60.80 \\
    \hline
    \textit{hoc-1} & \textbf{50.04} & 55.61 & 52.70 & 52.15 & 60.26 & 62.37 \\
    \hline
    \textit{hoc-2} & 64.91 & 56.40 & 63.65 & \textbf{37.09} & 67.05 & 59.00 \\
    \hline
    \hline
    avg. & 57.59 & 54.11 & 61.46 & \textbf{46.99} & 67.88 & 64.11 \\
    \hline
    \hline
    \textit{bb-3} & 60.41 & \textbf{33.94} & 59.22 & 42.64 & 60.61 & 55.56 \\
    \hline
    \textit{got-3} & 74.71 & \textbf{49.31} & 70.34 & 63.17 & 61.33 & 52.89 \\
    \hline
    \textit{hoc-3} & \textbf{57.68} & 59.87 & 67.52 & 67.41 & 70.55 & 67.05 \\
    \hline
    \hline
    avg. & 64.13 & \textbf{47.71} & 65.69 & 57.74 & 64.16 & 58.50 \\
    \hline
  \end{tabular}
\end{table}

Results are given both in taking as input the local speakers
hypothesized in each dialogue scene during the previous step
(\textit{input auto.}) as well as the real speakers manually annotated
(denoted \textit{input ref.}). In both cases, the second step of
clustering is performed in an unconstrained way (denoted~\textit{2S}),
allowing any local speakers to be clustered during the agglomerative
process, and in a constrained way (denoted~\textit{cst. 2S}), by
preventing it. For the sake of comparison, the results of two standard
speaker diarization tools (denoted \textsc{lia}, described
in~\cite{bozonnet2010lia}, and \textsc{lium}, described
in~\cite{meignier2010lium} and~\cite{rouvier2013open}), are also
reported: these tools receive in input all the spoken segments covered
by the dialogue patterns.

Though still high, the \textsc{der} is generally reduced by
integrating to the clustering process the structural information based
on visual patterns. By stopping the clustering before all the
instances can be gathered, the ``different-speakers'' property allows
to cut the resulting dendogram at a suitable level, providing an early
stop condition of the process, when only a few mutually exclusive
groups of instances remain. By contrast, unconstrained clustering has
to face the critical issue of finding the optimal partition of the
instances.

Table~\ref{nspk} reports the average number of speakers involved in
the dialogue scenes considered, as hypothesized by the different
systems.

\begin{table}[h]
  \caption{\label{nspk}{\it \textsc{der} Average number of
      hypothesized speakers}}
  \vspace{2mm}
  \centering
  \begin{tabular}{|c|c||cc||cc|}
    \hline
    & \textbf{truth} & \textit{2S} & \textit{cst. 2S} & \textsc{lia} & \textsc{lium}\\
    \hline
    \hline
    \textit{bb} & \textbf{10.3} & 7.3 & \textbf{11} & 6 & 25.7 \\
    \hline
    \textit{got} & \textbf{25.3} & 4.7 & 15.7 & 9.3 & \textbf{24} \\
    \hline
    \textit{hoc} & \textbf{20.7} & 3.7 & \textbf{24} & 6 & 27 \\
    \hline
  \end{tabular}
\end{table}

As can be seen, two systems (unconstrained 2-step clustering and
\textsc{lia}), tend to cut the clustering dendogram at a high level,
resulting in a few number of too wide classes. Conversely,
\textsc{lium}, by cutting the tree at a low level, overestimates in
two cases the number of speakers. The constrained clustering approach
(\textit{cst. 2S}), resulting in disjoint dendograms, offers an
reasonable approximation of the number of speakers and prevents early
as well as late cuts of the clustering tree.

\section{Conclusion and perspectives}
\label{sec:conclusion}

In this paper, we proposed to achieve speaker diarization of narrative
movies by relying on the structural information they carry. By
detecting similar shots, some covering patterns, typical of dialogue
scenes, can be extracted and a first step of speaker diarization can
be locally performed inside each dialogue boundaries. A second step of
clustering, aiming at detecting the recurring speakers, is then
applied to the locally hypothesized speakers: at each iteration of
this global clustering process, the constraint that speakers locally
assumed to be different must not be clustered is propagated; as a
result, the agglomerative process is blocked far before all the
instances are clustered, allowing a more convenient partition of the
initial set than when applying an unconstrained approach.

Despite the coverage of the visual patterns, there still remains some
sparse spoken segments outside their boundaries (near than a half of
the total amount of speech). A specific study of the shot patterns
involved in the dialogue scenes could allow to increase their
coverage. The labelling of the remaining spoken segments could then be
achieved by assigning them to the --~possibly noisy~-- speaker
models resulting from the \textsc{sd} process. Finally, visual
information could be used during the local clustering of the dialogue
utterances by exploiting the way the shots alternate with each other.

% References should be produced using the bibtex program from suitable
% BiBTeX files (here: strings, refs, manuals). The IEEEbib.bst bibliography
% style file from IEEE produces unsorted bibliography list.
% -------------------------------------------------------------------------
\bibliographystyle{IEEEbib}
\bibliography{refs}

\end{document}